\documentclass[12pt]{iopart}
\usepackage{bm}
\usepackage{epsf}
\usepackage{graphicx}
\begin{document}

\title[]{Quantum evolution of the Universe from $\tau=0$ in the constrained
quasi-Heisenberg picture}

\author{S.L. Cherkas\dag   \ and V.L. Kalashnikov\ddag
}

\address{\dag\
Institute of Nuclear Problems, Bobruiskaya 11, Minsk 220050,
Belarus}

\address{\ddag\ Technische Universit\"{a}t Wien, Gusshausstrasse 27/387, Vienna A-1040, Austria}

\begin{abstract}
The Heisenberg picture of the minisuperspace model is considered.
The  suggested quantization scheme interprets all the observables
including the Universe scale factor as the (quasi)Heisenberg
operators. The operators arise as a result of the re-quantization
of the Heisenberg operators that is required to obtain the
hermitian theory. It is shown that the DeWitt constraint $H=0$ on
the physical states of the Universe does not prevent a
time-evolution of the (quasi)Heisenberg operators and their mean
values. Mean value of an observable, which is singular in a
classical theory, is also singular in a quantum case. The
(quasi)Heisenberg operator equations are solved in an analytical
form in a first order on the interaction constant for the
quadratic inflationary potential. Operator solutions are used to
evaluate the observables mean values and dispersions. A late stage
of the inflation is considered numerically in the framework of the
Wigner-Weyl phase-space formalism. It is found that the
dispersions of the observables do not vanish at the inflation end.
\end{abstract}

\pacs{F06.60.Ds, 98.80.Hw, 98.80.Cq}


\maketitle
\section{Introduction}

A focus of the article is the construction of the quantization
scheme in which the observables (including Universe scale factor)
are the time dependent operators. This allows to evaluate their
mean values and dispersions.

A variety of the quantization schemes for the minisuperspace model
can be roughly divided in two classes: imposing the constraint
"before quantization" \cite{hen} and "after
quantization"\cite{wheel}. In the former the constraints are used
to exclude the "nonphysical" degrees of freedom. This allows then
to construct the Hamiltonian acting in the reduced "physical"
phase space.

The last schemes prefer imposing the constraint "after
quantization". This leads to the Wheeler-DeWitt equation on
quantum states of the Universe \cite{wheel,witt}. Our scheme can
be considered among the last class because we use the
Wheeler-DeWitt equation. However we supplement this equation with
the system of the quasi-Heisenberg operators acting in the space
of the solutions of this equation. Quantization rules for these
operators are defined consistently with the choice of the
hyperplane used for normalization of the solutions of the
Wheeler-deWitt equation in the Klein-Gordon style.

  In the simplest case of an isotropic and uniform Universe filled
with a scalar field, the minisuperspace equation contains only two
variables: the scale factor of Universe and the amplitude of the
scalar field. There is no an explicit "time" in the corresponding
Wheeler-deWitt equation, whereas we are interested namely in the
Universe evolution in time. This leads to various discussions
about "time disappearance" and interpretation of the wave function
of Universe \cite{hall}. Possible solutions like to introduce time
along the quasi-classical trajectories, or subdivide Universe into
classical and quantum parts were offered \cite{vil}. Such point of
view can not be satisfactory. Ideally, time does exist
independently regardless of whether we consider Universe quantum
or classically. That forces to consider the quantum evolution of
Universe without any need in foreign ingredients
\cite{kag,mil,geor,weist}. Let us remind that this situation is
analogous to that in string theory, where the constraint $H=0$
also exists. Nevertheless this constraint does not prevent an
evolution of the Heisenberg operators $\hat X(\sigma,\tau)$ along
$\tau$ \cite{kaku}.

In fact the constraint $H=0$ (in a theory of the constrained
systems $H$ is usually refereed as the super-Hamiltonian)  tells
nothing about whether the mean values of the Heisenberg operators
evolve with time or not. This is defined by the normalization of
the wave function. We can outline an issue in the following way.
Let there is the wave function $\psi(a,x)$ dependent on two
variables and obeying $\hat H\psi(a,x)=0$.

For evolution of some Heisenberg operator $\hat H$ we have
\begin{equation}
<A(\tau)>=<\psi|e^{i\hat H\tau}\hat A e^{-i\hat H\tau} |\psi>.
\end{equation}
At first sight it seems that there is no evolution. However this
is not the case. Really $e^{-i\hat H\tau} |\psi>=|\psi>$, but we
cannot write $<\psi|e^{i\hat H\tau}=<\psi|$. The point is that the
wave function can not be normalized in the ordinary way $\int
\psi^*(a,x)\psi(a,x)dxda=1$ due to the constraint. In fact the
function is unbounded along one of the variables. For instance,
let this is $a$ variable. So if $H$ contains differential operator like $\frac{%
\partial^2}{\partial a^2 }$, one cannot move $\frac{\partial^2}{\partial a^2}$
to the left by the habitual operation $<\psi|\frac{\partial^2}{\partial a^2 }=<%
\frac{\partial^2}{\partial a^2 }\psi|$ through the integration by
parts. As a result, the assumption that $<\psi|\hat H=<\hat
H\psi|=0$ is wrong.

Unfortunately this is an oversimplified picture. Reality was found
to be more complicated. In the next section we show that the
normalization of the wave function in the Klein-Gordon style
permits evolution of the mean values of the Heisenberg operators.
However, additional  efforts are needed to obtain a genuine
hermitian theory.

\section{Quantization of a particle-clock}

The problem of the quantum cosmology has many common features with
those of the relativistic particle \cite{hen,kaku,gitm,mori}. The
action for the relativistic particle can be defined as:

\begin{equation}
S=-m\int\sqrt{-\dot{x}_\mu^2}d\tau .  \label{s1}
\end{equation}

The equivalent form leading to Eq. (\ref{s1}) by means of
variation of the lapse function $e(\tau)$ is

\begin{equation}
S=\frac{1}{2}\int(e^{-1}\dot{x}_\mu^2-e \,m^2)d\tau,  \label{s2}
\end{equation}
where $(-,+,+,+)$ signature is used. One more equivalent form resulting in (%
\ref{s2}) due to varying $p_\mu$ looks as
\begin{equation}
S=\int\bigl\{p_\mu \dot{x}^\mu-e\left(\frac{p_\mu^2+m^2}{2}\right)\bigr\}%
d\tau.
\end{equation}

Using the re-parameterization invariance \cite{kaku,gitm} we can
choose the lapse function as $e=1/m$. One can see from the last
equation that the Hamiltonian is equal to
$H=\frac{p_\mu^2+m^2}{2m}$ and vanishes on the constraint surface
$H=0$ after varying on $e$. The quantization procedure is based on
the assumptions:
\[
[\hat p^\mu,\hat x^\nu]=-ig^{\mu\nu}, ~~~~\hat p^\mu=\{\hat
\varepsilon, \hat{ \bm p} \}\equiv\{i\frac{\partial}{\partial
t},-i\bm \nabla\}.
\]
Then the constraint becomes the Klein-Gordon equation. The
commutators of the position and four-momentum operators with the
Hamiltonian result in the Heisenberg equation of motion:
\begin{equation}
\frac{d \hat x^\mu}{d\tau}=i[\hat H,\hat x^\mu]=\frac{\hat
p^\mu}{m}, ~~~~~~ \frac{d\hat p^\mu}{d \tau}=0. \label{eqgeiz}
\end{equation}

\noindent Evident solution of the motion equation is
\[
\hat x^\mu(\tau)=x^\mu+\frac{p^\mu(0)}{m}\,\tau, ~~~\hat
p^\mu(\tau)=\hat p^\mu(0),
\]
where $\hat x^{\mu}(0)=x^{\mu}\equiv\{t,\bm r\}$ and $\hat
p^\mu(0)=\hat p^\mu$. Because before quantization we do not use
any gauge fixing providing the time-dependent second class
constraint \cite{gitm}, $\hat t(\tau)$ can be treated as an
operator. We shall refer to $\hat t(\tau)$ as "physical time" and
to $\tau$ as "proper time". Whereas the physical time is operator,
the proper time is some parameter "always running forward" like
time in the Newtonian physics.

One can check that the wave function satisfying the Klein-Gordon
equation cannot be normalized through the integration over $d^4x$
\cite{vil}. However, it can be normalized through the time-like
component of the conserved four-current:
\[
<\psi|\psi>=-i\int \{\psi(\bm r,t)\partial_t\psi^*(\bm
r,t)-(\partial_t\psi(\bm r,t))\psi^*(\bm r, t)\}d^3\bm r.
\]

\noindent Normalized wave packets have the form:
\[
\psi(x)=\int \frac {a(\bm k)e^{-i\varepsilon(\bm k)t+i\bm k\bm r}} {\sqrt{%
2\varepsilon(\bm k)(2\pi)^3}}d^3\bm k;~~~\int \mid a(\bm
k)\mid^2d^3\bm k=1,
\]
where $\varepsilon^2(\bm k)=\bm k^2+m^2$. To avoid an appearance
of the states with the negative norm we have to choose only
positive frequency solutions.  For this wave packet the
integration $\int \psi^*(x)\psi(x)d^4x$ does
not converge. Hence the wave function is unbounded along the $%
t$-variable.

Let us define the mean value of some Heisenberg operator $\hat
A(\tau)$ for the case of a free relativistic particle as
\begin{eqnarray}
<\hat A(\tau)>=\frac{-i}{2}\int \biggl( \psi^*(\bm r,t)\hat A(\bm
r,t,\hat \varepsilon,\hat {\bm p},\tau)\partial_t\psi(\bm r,t)\nonumber\\
~~~~~~~~~~~~~~~~~~~~~~~~~~ - (\partial_t\psi^*(\bm r,t))\hat A(\bm
r,t,\hat
\varepsilon,\hat {\bm p},\tau)\psi(\bm r,t)\biggr)d^3\bm r \biggr| %
_{t=0}+h.c.,  \label{mean_1}
\end{eqnarray}
where $"h.c."$ means the hermitian conjugate value.

The definition (\ref{mean_1}) has the following properties:

\noindent  1) It is consistent with the normalization of the wave
function if we choose $\hat A$ to be equal to the unit operator.

\noindent 2) It looks like as an expression for the mean value of
the Heisenberg operator in the nonrelativistic picture when the
operator acts on the wave function taken at the initial moment of
time (note, that after integration over $d^3\bm r$ we have to set
$t=0$).

\noindent 3) It has a natural property $\frac{<\hat A(\tau)>}{d\tau}=<\frac{%
\hat A(\tau)}{d\tau}>$.

\noindent 4) It gives zero for the physical time, when the proper
time is equal to zero.

\begin{figure}[h]
\includegraphics{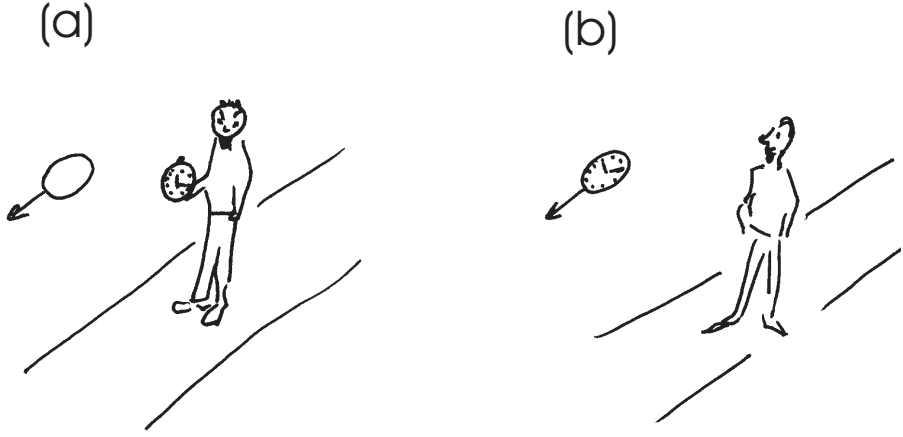}
\caption{\label{stoks}(a) Observer describing a particle motion by
own clock, (b) Observer describing a particle motion by "particle
clock".} \label{stoks}
\end{figure}

Let us emphasize the difference between our quantization scheme
for the relativistic particle and e.g. the quantization schemes
(in the ADM style with the some gauge fixing or in the
Foldy-Wouthauthen representation of the Klein-Gordon equation)
where the time is not an operator. Let consider some observer with
an "exact" clock tracking a particle motion. The observer can
measure, for instance, the particle coordinate and describe it by
the operator $\hat {\bm r}(t)$ to take into account the quantum
fluctuations. There remain only technical problems like to find
the Hamiltonian translating $\hat {\bm r}(t)=e^{i\hat H t}\hat{\bm
r}(0) e^{-i\hat H t}$ etc. Let consider another situation, namely
the observer has no a clock but there is an exact "clock" on the
particle (e.g., this is the "radioactive" particle emitting
photons every equal time pieces in a reference system connected
with the particle). The observer having no own clock is forced to
measure time by the particle clock. The measured time is the
fluctuating value described by the operator $\hat
t(\tau)=\frac{\hat \varepsilon}{m}\tau$, because the particle
energy $\varepsilon$ is the fluctuating quantity for the wave
packet. Thus our quantization scheme describes namely the particle
clock (i.e. the particle supplied with the clock) rather than the
"mute" particle.

After averaging, the Heisenberg equations of motion result in
\begin{equation}
<\hat t(\tau)>=<\hat \varepsilon> \frac{\tau}{m},~~ <\hat {\bm
r}(\tau)>=<{\bm r}>+ <\hat {\bm p}>\frac{\tau}{m}.
\end{equation}
One can see that the physical time is proportional to the proper
time. It is interesting to calculate the dispersion of the
physical time for the
Gaussian wave packet $a(\bm k)\sim e^{-\alpha\bm k^2}$ with the $\frac{1}{%
\alpha}>>m^2$. An evaluation of the mean values of $\hat \varepsilon =i%
\frac{\partial }{\partial t}$ and its square gives:
\[
<\hat \varepsilon>=\frac{\int_0^\infty \varepsilon (k)e^{-\alpha k^2 }k^2dk}{%
\int_0^\infty e^{-\alpha k^2
}k^2dk}\approx\frac{2}{\sqrt{\pi\alpha}},~~<\hat
\varepsilon^2>=\frac{\int_0^\infty \varepsilon^2 (k)e^{-\alpha k^2
}k^2dk}{\int_0^\infty e^{-\alpha k^2
}k^2dk}\approx\frac{3}{2\alpha},
\]
that results in
\begin{eqnarray*}
\frac{\sqrt{<\hat t^2>-<\hat t>^2}}{<\hat t>}=\frac{\sqrt{<\hat
\varepsilon^2>-<\hat \varepsilon>^2}}{<\hat \varepsilon>}=\sqrt{\frac{3\pi}{8%
}-1}.
\end{eqnarray*}
Thus, the "particle-clock" is a bad clock when the particle is
localized in the spatial region less than the Compton wave length.
Let us imagine what occurs when the particle possessing the
electric charge is placed in the electric field. Let there is a
cloud of the particles with some initial dispersion of the energy
placed in the electric field. The particles begin to accelerate in
the electric field and thereby they receive the energies much
larger than their initial energies. As a result the relative
energy dispersion becomes negligible. Similar picture holds for
the quantum case so that the relative accuracy of the
"particle-clock" increases.

Let us remind that in fact we have used here the covariant proper
time formalism, which goes down to Fock \cite{fock} and Kramers
\cite{kram}. This formalism is widely used for derivation of the
Bargman-Michel-Telegdi equations for the particle spin motion in
the external fields \cite{yam}.  In this formalism we have to
normalize the wave function through the Klein-Gordon scalar
product instead of normalization through the integration over
$d^4x$ (on the other side this permits the time-evolution despite
of $\hat H\psi=0$). Absence of hermicity leads to the complex mean
value of some observables. Although we are try to avoid a
complexity by adding "$h.c.$" quantity in Eq. (\ref{mean_1}) the
negative dispersion appears when we evaluate, for instance, $<\hat
t^4(\tau)>$ etc. Thus we have seen that the imposing constraint to
the state vectors is not sufficient to obtain a "good" theory in
the framework of the covariant proper time formalism. In next
section we shall propose solution for this problem.

\section{Quantum cosmology}

Let us start from the Einstein action of a gravity and the action
for an one-component real scalar field:
\begin{equation}
S=\frac{1}{16\pi G}\int d^4 x\sqrt{-g}R+\int d^4 x\sqrt-g[\frac{1}{2}%
(\partial_\mu\phi)^2-V(\phi)],
\end{equation}
where $R$ is the scalar curvature and $V$ is the matter potential
which includes a possible cosmological constant effectively. We
restrict our consideration to the homogeneous and isotropic
metric:
\begin{equation}
ds^2=N^2(\tau)d\tau^2-a^2(\tau)d\sigma^2.
\end{equation}
Here the lapse function  $N$ represents the general time
coordinate transformation  freedom. For the restricted metric the
total action becomes
\[
S=\int N(\tau)\biggl\{ \frac{3}{8\pi G}a\left( {\mathcal K
}-\frac{\dot{a}^2}{N^2(\tau)} \right)
+\frac{1}{2}a^3\frac{\dot{\phi}^2}{N^2(\tau)}-
a^3V(\phi)\biggr\}d\tau,
\]
where $\mathcal{K}$ is the signature of the spatial curvature.
This action can be obtained from the following expression by
varying on $p_a$ and $p_\phi $
\[
S=\int\biggl\{p_\phi\dot{\phi}+p_a\dot{a}-N(\tau) \biggl(-\frac{3a{\mathcal K}}{8\pi G}%
-\frac{8\pi G\,p_a^2}{12a}
+\frac{p_\phi^2}{2a^3}+a^3V(\phi)\biggr) \biggr\}d\tau.
\]
Varying on $N$ gives the primary constraint
\begin{eqnarray}
H=-\frac{3a{\mathcal K}}{8\pi G}-\frac{8\pi G\,p_a^2}{12a}+\frac{p_\phi^2}{2a^3}%
+a^3V(\phi)=0. \label{constr1}
\end{eqnarray}
After quantization $[\hat a,\hat p_a]=-i$, $[\hat \phi,\hat
p_\phi]=i$ this constraint turns into the DeWitt equation
$
\hat H\psi(a,\phi)=0.
$
Looking at the constraint equation a desire may appear to
modernize or remove it \cite{lasuk}. Apparently this implies (both
on classical and on quantum levels) existence of some preferred
system of reference. Although there are some logically consistent
theories implying preferred system of reference, for instance, the
Logunov relativistic theory of gravity \cite{log} giving an
adequate description of the Universe expansion \cite{kalash}, we
shall keep to the General Relativity here and retain the
constraint.

Let us first consider the flat Universe (${\mathcal K }=0$) with
$V(\phi)=0$ (corresponding Hamiltonian is $\hat H_0$).

Procedure, which is invariant under a general coordinate
transformation consists in postulating the quantum Hamiltonian
\cite{witt1}
\begin{equation}
\hat H_0=\frac{1}{2}g^{-\frac{1}{4}}\hat p_\mu g^{\frac{1}{2}}\hat
g^{\mu\nu}\hat p_\nu g^{-\frac{1}{4}},
\end{equation}
where $\hat p_\mu=-i g^{-\frac{1}{4}}\, \frac{\partial}{\partial x^\mu}\, g^{%
\frac{1}{4}}=-i\left(\frac{\partial}{\partial x^\mu}+\frac{1}{4}(\frac{%
\partial \ln g}{\partial x^\mu})\right)$. For our choice of variables $%
x^{\mu}=\{a,\phi\}$, $p_\mu=\{-p_a,p_\phi\}$ metric has the form
(in the units $4\pi G/3=1$)
$
g^{\mu\nu}\ = \left(
\begin{array}{cc}
-\frac{1}{a}\; & \;0 \\
\, & \, \\
\,\,0\; & \;\frac{1}{a^3}%
\end{array}
\right), $
so that $g=a^4$, $\hat p_a=i\left(\frac{\partial}{\partial a }+\frac{1}{a}%
\right)$ and the Hamiltonian is
\begin{equation}
\hat H_0=-\frac{1}{4}\left(\hat p_a^2\frac{1}{a}+\frac{1}{a}\hat
p_a^2\right)+\frac{\hat p_\phi^2}{2 a^3}= \frac{1}{2 a^2}\frac{\partial}{%
\partial a} a\frac{\partial}{\partial a}-\frac{1}{2 a^3}\frac{\partial^2}{%
\partial \phi^2}.
\end{equation}

Explicit expression for the wave function satisfying $\hat
H_0\psi=0$ is
\[
\psi_k(a,\phi)=a^{\pm i|k|}e^{ik\phi}.
\]

Exactly as in the case of the Klein-Gordon equation we should
choose only the positive frequency solutions \cite{vil}. Such wave
function corresponds to the definite choice of the boundary
condition for the minisuperspace: the wave function is formed by
the modes bounded on $k$ and only ingoing in a singularity.

Thus the wave packet
\begin{equation}
\psi(a,\phi)= \int c(k)\frac{a^{-
i|k|}}{\sqrt{4\pi|k|}}e^{ik\phi}d k \label{pack} \label{ff}
\end{equation}
will be normalized by
\begin{equation}
ia\int\left(\frac{\partial \psi}{\partial a}\psi^*-\frac{\partial \psi^*}{%
\partial a}\psi \right)d\phi=\int c^*(k)c(k)dk=1.  \label{nm}
\end{equation}

The proper time evolution of the operators
\[
\frac{d\hat A(\tau)}{d\tau}=i[\hat H,\hat A]
\]
results in
\begin{equation}
\fl~{\hat p}_\phi^{\bm\cdot}(\tau)=0,~~{\hat p}_\phi(\tau)=const,
~~(\hat a^3(\tau))^{\bm\cdot}=\frac{3}{2}\left(\hat p_a a+a \hat
p_a\right), ~~ \frac{3}{2}(\hat p_a a+ a \hat
p_a)^{\bm\cdot}=-9\hat H_0. \label{eq}
\end{equation}

Set of equations for the Heisenberg operators and the constraint
equation for the states would be considered as a tool to evaluate
the mean values of observables. The obstacle is that the operators
are hermite relatively an integration over $\sqrt{g}d^n x$, while
due to constraint the states can not be normalized in such a way
and are normalized in the Klein-Gordon style.

Our recept consists in enforcing the constraints on the equation
of motion for the Heisenberg operators at $\tau=0$. First let us
remind the Dirac quantization procedure \cite{dirac} and return to
the classical picture for
this goal. According to Dirac, besides the primary constraint $\Phi_1=-\frac{%
p_a^2}{a}+\frac{p_\phi^2}{a^3}$ (see (\ref{constr1})), we have to
set some additional gauge fixing (secondary) constraint, which can
be chosen in our case as $\Phi_2=a-const$, because the hyperplane
$a=const$ is chosen earlier for the normalization of the wave
function in the Klein-Gordon style. In contrast to the usual
formalism \cite{hen,gitm,tyu,shab}) we impose
constraints $\Phi_1=0$ and $%
\Phi_2=0$, which have to be satisfied only at hyperplane $\tau=0$.
Besides the ordinary Poisson brackets
\begin{equation}
\{A,B\}=\frac{\partial A}{\partial p_\mu}\frac{\partial B}{\partial x^\mu}-%
\frac{\partial A}{\partial x^\mu}\frac{\partial B}{\partial
p_\mu},
\end{equation}
the Dirac brackets are introduced
\begin{equation}
\{A,B\}_D=\{A,B\}-\{A,\Phi_i\}(C^{-1})_{ij}\{\Phi_j,B\},
\end{equation}

\noindent where $C$ is the nonsingular matrix with the elements $%
C_{ij}=\{\Phi_i,\Phi_j\}$ and $C^{-1}$ is the inverse matrix.
Quantization consists in the substitution of the commutators in
the Dirac brackets and the replacement of the variables by the
operators:
\begin{equation}
[\hat{\bm\eta},\hat{\bm\eta}^\prime]=-i\{\bm \eta, \bm \eta^\prime \}_D%
\biggr|_{\bm \eta\rightarrow\hat{\bm \eta}}.
\end{equation}
Here $\bm \eta$ implies the set of the canonical variables $p_\mu,
x^\nu$.

The Heisenberg equations of motion have to satisfy the constraints
at initial moment $\tau=0$ and the operators obey the commutation
relations obtained from the Dirac quantization procedure. Direct
evaluation gives
\begin{eqnarray}
[{\hat p}_a(0),{\hat a}(0)]=0,~~~ {[\hat p_\phi(0),\hat
a(0)]=0},\nonumber\\
 {[{\hat p}_\phi(0),{\hat \phi}(0)]=-i},~
{[{\hat p}_a(0),~\hat \phi(0)]=-i\frac{{\hat p}_\phi(0)}{{\hat p}_a(0){\hat a}%
^2(0) }}.  \label{rec0}
\end{eqnarray}

We have to solve the equations (\ref{eq}) with the given initial
commutation relations. This changes the canonical commutation
relations between the Heisenberg operators at $\tau\ne 0$.
Therefore we shall call theirs as the quasi-Heisenberg operators.
In fact we appeal to the structure of the classical theory and
re-quantize the equations of motion.

One can see that the commutation relations (\ref{rec0}) can be
satisfied through
\begin{equation}
\hat a(0)=const=a, ~~\hat p_\phi(0)=\hat p_\phi, ~~\hat
p_a(0)=|\hat p_\phi|/a , ~~\hat \phi(0)=\phi,  \label{rec1}
\end{equation}
where $\hat p_\phi=-i\frac{\partial}{\partial \phi}$. Variable $a$ is $c$%
-number now because it commutes with all operators \cite{shab}.
Solutions of the equations (\ref{eq}) are
\begin{eqnarray}
\hat p_\phi(\tau)=\hat p_\phi,~~~~\hat \phi(\tau)=\phi+\frac{\hat
p_\phi}{3|\hat p_\phi|}\ln(a^3+3|\hat p_\phi|\tau)-\frac{\hat
p_\phi}{|\hat
p_\phi|}\ln\,a,  \nonumber \\
\hat p_a(\tau)=\frac{|\hat p_\phi|}{(a^3+3|\hat
p_\phi|\tau)^{1/3}},~~~~\hat a^3(\tau)=a^3+3|\hat p_\phi|\tau,  \nonumber \\
\end{eqnarray}

We imply that these quasi-Heisenberg operators act in the Hilbert
space with the Klein-Gordon scalar product.

Mean value of some operator $\hat A(\tau)$ can be defined  as
\cite{groot}:
\begin{eqnarray}
<\hat A(\tau)>=ia\int\biggl(\psi^*(a,\phi)\hat A(\tau,\phi,\hat
p_\phi,a )\frac{\partial \psi(a,\phi)}{\partial a}~~~~~  \nonumber \\
-\frac{\partial \psi^*(a,\phi)}{\partial a}\hat A(\tau,\phi,\hat p_\phi,a )\psi(a,\phi) \biggr)%
d\phi\biggr|_{a\rightarrow 0}.~~  \label{nmm0}
\end{eqnarray}
\noindent However we shall use another definition:
\begin{eqnarray}<\hat A(\tau)>=i\,a\int \biggl( \psi^*(a,\phi)
D^\frac{1}{4} \hat A(\tau)D^{-\frac{1}{4}}
\frac{\partial}{\partial a} \psi(a,\phi)~~\nonumber
\\
- \frac{\partial}{\partial a }\psi^*(a,\phi)D^{-\frac{1}{4}}\hat
A(\tau)D^{\frac{1}{4}}\psi(a,\phi)\biggr)d\phi \biggr|
_{a\rightarrow 0},~~ \label{nmm}
\end{eqnarray}
where operator $D=-\frac{\partial^2}{\partial
\phi^2}+2\,a^6V(\phi)$ (since $a\rightarrow 0$ the $V$-term can be
omitted in the expression for $D$). The mean value (\ref{nmm}) is
particular case of that suggested in Ref. \cite{ali}, where an
one-particle picture of the Klein-Gordon equation in the
Foldy-Wouthausen representation is considered. The advantage of
this normalization can been seen in the momentum
representation of $\phi$ variable, where $\hat p_\phi=k$ and $%
\hat \phi= i\frac{\partial}{\partial k}$. Eq. (\ref{nmm0}) gives
\begin{eqnarray*}
 <\hat A(\tau)>=\frac{1}{2}\int {a}^{i|k|}
 \sqrt{|k|}c^*(k)\hat A(\tau,\hat \phi,k,a)a^{-i|k|}{{c(k)}\over{\sqrt{|k|}}}
 dk\nonumber\\+
\frac{1}{2}\int {a}^{i|k|}{{c^*(k)}\over{\sqrt{|k|}}}\hat
A(\tau,\hat
\phi,k,a)a^{-i|k|}\sqrt{|k|}c(k)dk\biggr|_{a\rightarrow 0},~~
\end{eqnarray*}
while Eq. (\ref{nmm}) leads to
\begin{eqnarray}
 <\hat A(\tau)>=\int {a}^{i|k|}c^*(k)\hat A(\tau,\hat \phi,k,a){a}^{-i|k|}c(k)dk\biggr|_{a\rightarrow
0}, \label{defali}
\end{eqnarray}
which is similar to the ordinary quantum mechanics and certainly
posses hermicity.

 Evaluation of the mean value over the wave packet
gives
\begin{equation}
<a^3(\tau)>=3\tau\int|k||c(k)|^2 dk.
\end{equation}
If we assume that there is no external "physical time" in Universe
(proper time $\tau$ hardly can be considered as the measurable
observable), then we must take some quantity, for instance,
$<a^3(\tau)>$ as the "physical time".

Next interesting quantity is the mean value of the scalar field
$<\hat\phi(\tau)>$:

\begin{eqnarray}
 \fl<\hat\phi(\tau)>_{a}=\int {a}^{i|k|}c^*(k)\left(
i\frac{\partial}{\partial k}+\frac{k}{3|k|}\ln\left( a^3+3|k|\tau
\right)-\frac{k}{|k|}\ln a \right)a^{-i|k|}c(k) dk
\nonumber\\
=\int\left( c^*(k)i\frac{\partial}{\partial
k}c(k)+\frac{k}{3|k|}\ln\left(a^3+3|k|\tau\right)
|c(k)|^2\right)dk. \label{canc}
\end{eqnarray}

Brackets $<\dots>_a$ with the index $a$ in (\ref{canc}) mean that
we do not set $a$ equal to zero yet (compare with Eq.
(\ref{defali})). A
remarkable property of Eq. (\ref{canc}) is that the term $-\frac{k}{|k|%
}\ln a$ cancels the term arising from the differentiation: $a^{i|k|}i\frac{%
\partial}{\partial k}a^{-i|k|}= \frac{k}{|k|}\ln a$. Thus we may get to $%
a\rightarrow 0 $ and obtain
\begin{equation}
<\hat \phi(\tau) >=\int\left(\frac{k}{3|k|}\ln(3|k|\tau)|c(k)|^2 +
c^*(k)i \frac{\partial}{\partial k}c(k)\right)dk.  \label{phi_1}
\end{equation}
Cancellation of the terms divergent under $a\rightarrow0$ in the
mean values of the quasi-Heisenberg operators is a general feature
of the theory, and give us possibility to evaluate, for instance,
\begin{equation}
<\hat \phi^2(\tau)
>=\int\biggl(\frac{1}{9}\ln^2(3|k|\tau)|c(k)|^2-c^*(k)\frac{\partial^2}{\partial
k^2}c(k)\biggr)dk. \label{phi2}
\end{equation}

One should not confuse the divergence at $a\rightarrow 0$ arising
under evaluation of the mean values with the singularity at
$\tau\rightarrow 0$. The mean values of operators, which are
singular at $\tau\rightarrow0 $ in classical theory remain
singular also in the quantum case. The way to avoid a singularity
is to guess, that Universe was burn not from a point but from a
"seed" of $a_0$-"size". Then in the expression for mean value we
have to assume $a\rightarrow a_0$ instead of $a\rightarrow 0$.
This puts a question about underling theory giving size of the
seed.

One more kind of the infinity can be found in  Eqs. (\ref%
{phi_1}), (\ref{phi2}). For $c(k)$, which does not tend to zero at
small $k$ the mean values of $\phi(\tau)$ and $\phi^2(\tau)$
diverge. This is a manifestation of the well known infrared
divergency of the scalar field minimally coupled with gravity.
Thus not all possible $c(k)$ are suitable for the construction of
the wave packets.

Let us consider Hamiltonian, containing the cosmological constant
$V_0$:
\begin{equation}
\hat H=\hat H_0+a^3V_0.  \label{ham_v}
\end{equation}
Explicit solution for the wave function $H\psi=0$ has the form
\begin{equation}
\psi_k(a,\phi)=\left(\frac{18}{V_0}\right)^{\frac{i|k|}{6}} \Gamma(1-\frac{i|k|%
}{3})J_{-\frac{i|k|}{3}}(\frac{\sqrt{2V_0}}{3}a^3)e^{i k\phi},
\label{ww}
\end{equation}
where $\Gamma(z)$ is the Gamma function and $J_\mu(z)$ is the
Bessel function. The wave function (\ref{ww}) tends to
$a^{-i|k|}e^{i k\phi}(1+O(a^6))$ asymptotically under
$a\rightarrow0$. If we assume that the operators of all physical
quantities are local on the $a$ variable or can be approximated by
the local ones, then for an evaluation of the mean values we may
always build the wave packet from the  functions $a^{-i|k|}e^{i k
\phi }$. This supply ideology of the usual nonrelativistic
Heisenberg picture which says that all dynamics is contained in
the Heisenberg operators and only knowledge of the wave function
is required. The argumentation holds for any potential $V(\phi)$,
because it contributes into the Hamiltonian as a term multiplied
by $a^3$.

Equations of motion obtained from the Hamiltonian (\ref{ham_v})
are
\begin{eqnarray}
(\hat a^3(\tau))^{\bm\cdot}=\frac{3}{2}(\hat p_a a+a \hat p_a); \\
\frac{3}{2}(\hat p_a a+a \hat p_a)^{\bm\cdot}=9V_0a^3-9\hat H_0;
\label{a2}
\\
9(V_0a^3-\hat H_0)^{\bm\cdot}=18V_0\frac{3}{2} (\hat p_a a+ a \hat
p_a). \label{a3}
\end{eqnarray}
The additional term $a^3V_0$ does not change relations
(\ref{rec0}) required for the re-quantization procedure. Only
expression for $\hat p_a(0)$ changes in (\ref{rec1}): $\hat
p_a(0)=\sqrt{\frac{\hat p_\phi^2}{a^2}+2V_0a^4}$.

Finally we arrive to
\begin{equation}
\hat a^3(\tau)=a^3+3|\hat p_\phi|\frac{\sinh(\tau\sqrt{18 V_0})}{\sqrt{18 V_0%
}} + a^3(\cosh(\tau\sqrt{18 V_0})-1).  \label{os}
\end{equation}

\noindent Evaluation of the mean values according to Eq.
(\ref{nmm}) leads to
\begin{eqnarray*}
<\hat a^3(\tau)>=\frac{\sinh (3\,\sqrt{2V_0}\,\tau )}
{\sqrt{2V_0}}\int
|k||c(k)|^2dk, \\
<\hat a^6(\tau)>=\frac{\left({\sinh (3\,\sqrt{2V_0}\,\tau )}\right)^2}{2\,V_0%
}\int k^2 |c(k)|^2dk.
\end{eqnarray*}
This shows, that the dispersion $\frac{\sqrt{<\hat a^6>-<\hat
a^3>^2}}{<\hat a^3>}$ does not depend on $\tau$, exactly as in the
case of the free relativistic particle. Thus in this model the
evolution of Universe remains quantum during all time. This
results from the absence of any scale length (but not from an
ambiguity of the wave function normalization; see, for instance,
Ref. \cite{hab1}). Such a length appears if we take
$V(\phi)=\frac{m^2\phi^2}{2}$. One may suggest that the expansion
of Universe remains quantum until the moment, when the scaling
factor approaches the Compton wave length $1/m$. When $<a(\tau)>$
becomes greater than $1/m$ the expansion can be described
classically. To see this explicitly we have to find the
quasi-Heisenberg solutions with the above potential.

\section{Operator equations for the quadratic inflationary potential}
As it has been discussed the quantization procedure is reduced to
the quantization of the equations of motion i.e. considering  them
as the operator equations, which have to be solved with the
initial conditions satisfying to the constraint  at $\tau=0$. For
the Hamiltonian
\begin{equation}
H=-\frac{p_a^2}{2 a}+\frac{p_\phi^2}{2 a^3}+a^3g\frac{\phi^2}{2}
\end{equation}
we have the equations
\begin{eqnarray}
\ddot a=-\frac{3}{2}a\dot \phi^2-\frac{\dot a^2}{2 a}+\frac{3}{2}a
g \phi^2, \nonumber \\\ddot \phi=-3\frac{\dot a}{a}\dot \phi-g\phi
\label{infl}
\end{eqnarray}
and the constraint
\begin{equation}
-{\dot a^2}{a}+{\dot \phi^2}{a^3}+a^3g\phi^2=0. \label{conop}
\end{equation}

\begin{figure}
\begin{center}
\epsfbox{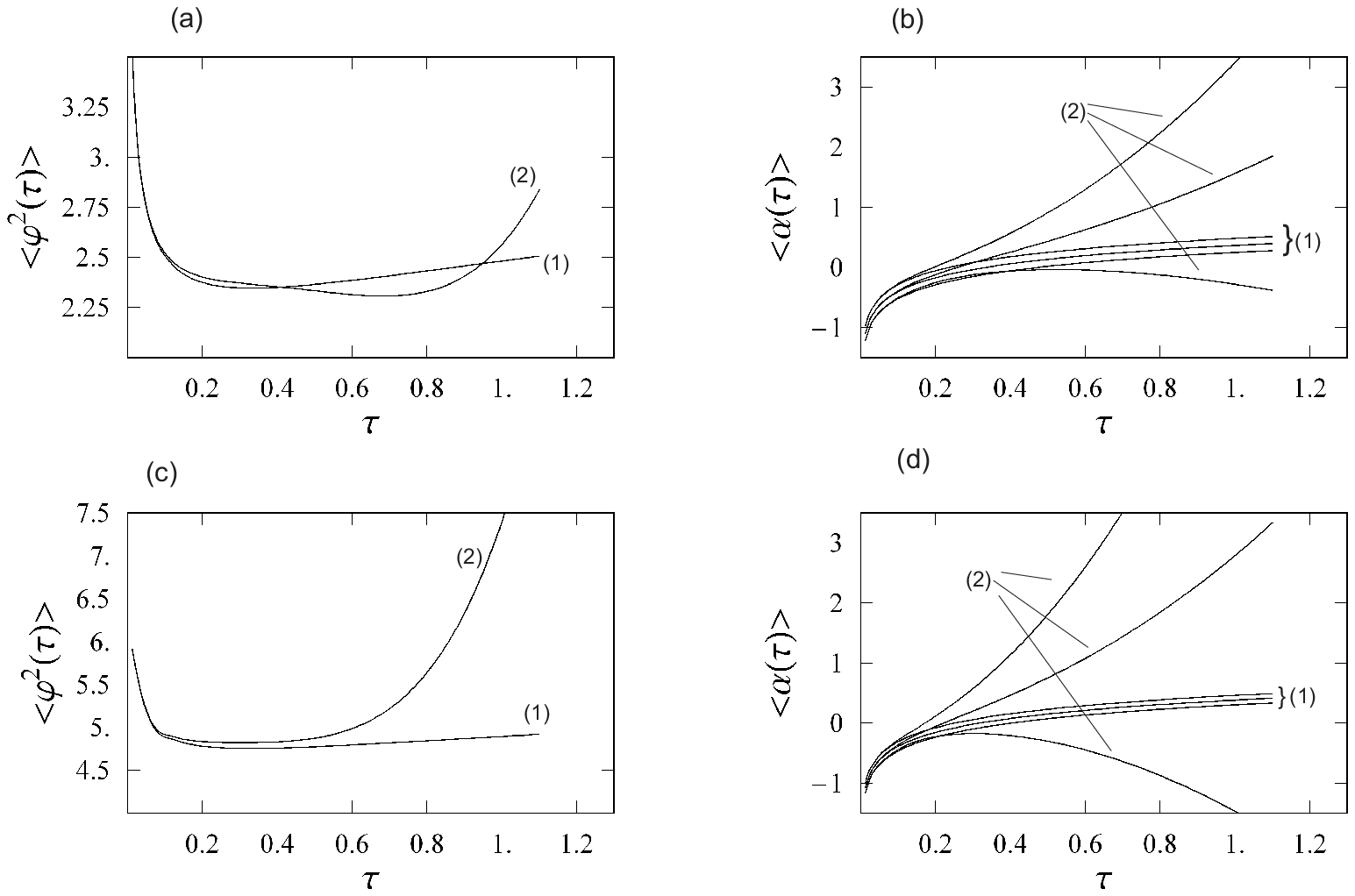}
\end{center}
 \vspace{0.5 cm}
\caption{\label{fig2} Mean values of the square of the scalar
field and the logarithm of the Universe scale factor calculated in
the zero (1) and first orders (2) on $g=1$, for the wave packet
$c(k)=\frac{4}{\sqrt{3}}\left(\frac{2}{\pi}\right)^{1/4}k^2\,\exp(-k^2)$
-- Fig. (a), (b), and for the wave packet
$c(k)=\left(\frac{2}{\pi}\right)^{1/4}\,\exp(-k^2-\frac{1}{k^2}+2)$
-- Fig. (c), (d). Each of the triplets of the curves in Fig. (b),
(d) corresponds to the $<\hat \alpha(\tau)>-\sqrt{D(\tau)}$
(lower), $<\hat \alpha(\tau)>$ (medial), $<\hat
\alpha(\tau)>+\sqrt{D(\tau)}$ (upper), where $D=<\hat
\alpha^2(\tau)>-<\hat \alpha(\tau)>^2$. }
 \vspace{0. cm}
\end{figure}

The point means the differentiation over $\tau$. After
quantization Eqs. (\ref{infl}) lead to the equations for the
quasi-Heisenberg operators, which have to be solved with the
operator initial conditions
\begin{eqnarray}
\hat a(0)\equiv a,~~~
 \hat \phi(0)\equiv\phi,~~~
\hat p_\phi(0)\equiv -i\frac{\partial}{\partial \phi},~~~ \dot
{\hat \phi}(0)=\frac{\hat p_\phi(0)}{\hat
a^3(0)}=\frac{1}{a^3}\left(-i\frac{\partial}{\partial \phi}\right), \nonumber\\
\dot {\hat a}(0)=\hat a(0)\sqrt{\dot {\hat \phi}^2(0)  +g {\hat
\phi}^2(0)} =\sqrt{\frac{1}{a^4}\left(-i\frac{\partial}{\partial
\phi }\right)^2+g a^2\phi^2}.
\end{eqnarray}
According to our ideology the operator constraint (\ref{conop}) is
satisfied only at $\tau=0$. The ordinary problem of the operator
ordering arises, because in the general case the quasi-Heisenberg
operators are noncommutative. The problem becomes more transparent
if we change the variable $\hat \alpha= \ln \hat a$:
\begin{eqnarray}
\ddot{\hat\alpha} +\frac{3}{2}\dot{\hat\alpha}^{2}-\frac{3}{2}
g\,\hat\phi^2+\frac{3}{2}{\dot{\hat\phi}}^{2}=0,\nonumber\\
{\ddot{\hat\phi}} +\frac{3}{2}\left(\dot{\hat
\alpha}\dot{\hat\phi}+\dot{\hat\phi}\dot{\hat\alpha}\right)+g\hat\phi=0.
\label{syst}
\end{eqnarray}
This system has to be solved with the initial conditions $\hat
\phi(0)\equiv\phi$, $\hat \alpha(0)=ln a$, $\hat {\dot
\phi}(0)=\frac{1}{a^3}\left(-i\frac{\partial}{\partial
\phi}\right)$, $ \dot{\hat
\alpha}(0)=\sqrt{g\phi^2+\frac{1}{a^6}\left(-i\frac{\partial}{\partial
\phi }\right)^2} $. We have used the symmetric ordering in Eq.
(\ref{syst}).

The operator equations under consideration can be solved within
the framework of the perturbation theory in the first order on
interaction constant. The solution in analytical form is given  in
Appendix. The calculations of the mean values based on
(\ref{defali}) have been carried out with the two kinds of the
wave packets:
$c(k)=\frac{4}{\sqrt{3}}\left(\frac{2}{\pi}\right)^{1/4}k^2\,\exp(-k^2)$
and
$c(k)=e^2\left(\frac{2}{\pi}\right)^{1/4}\,\exp(-k^2-\frac{1}{k^2})$,
which were normalized as $\int_\infty^\infty |c(k)|^2dk=1$.
Because of the infrared divergency we must take only wave packets
with vanishing $c(k)$ at $k\rightarrow 0$. The results of the
calculations are shown in Figs. \ref{fig2}. The packets under
consideration are symmetric thus the mean value of the scalar
field vanishes. In the first order on $g$ we can consider only an
early beginning of the inflation. At this stage the square of the
the scalar field grows due to initial "kinetic energy". The
logarithm of the scale factor grows linearly.

As it  has been discussed in the previous section in the
zero-order on $g$, the relative dispersion of $\hat a^3(\tau)$
does not depend on $\tau$ so that the dispersion of $\hat
\alpha(\tau)=\ln \,\hat a(\tau)$ is constant. In the first order
on $g$ we might expect that the dispersion of the scale factor
should decrease and the quantum universe will approach to the
classical one. However, we see an opposite picture: the dispersion
of $\alpha$ grows with $\tau$. However, as it is suggested in Ref.
\cite{hab2} some mechanism of the classical world appearance has
to exist.

\section{Wigner-Weyl evolution of the minisuperspace}

\noindent The analysis of the inflation at its late stages
requires a numerical consideration of the operator equations that
can be realized within the framework of the Weyl-Wigner
phase-space formalism \cite{groot}. Let us remind that in this
formalism every operator acting on $\varphi$ variable have the
Weyl symbol: ${\mathcal W }[\hat A]=A(k,\phi)$. The simplest Weyl
symbols reads: $\mathcal W[-i\frac{\partial }{\partial \phi}]=k$,
${\mathcal W[\phi]=\phi}$. Weyl symbol of the symmetrized product
of operators has the form
\[
\fl W[\frac{1}{2}(\hat A\hat B+\hat B\hat
A)]=\cos\left(\frac{\hbar}{2}\frac{\partial }{\partial
\phi_1}\frac{\partial }{\partial
k_2}-\frac{\hbar}{2}\frac{\partial }{\partial
\phi_2}\frac{\partial }{\partial
k_1}\right)A(k_1,\phi_1)B(k_2,\phi_2)\biggr|_{k_1=k_2=k,\,\phi_1=\phi_2=\phi}\,,
\]
 where the Plank constant is restored only to point up to which
 order we will expand the cosine in the next.
This has no direct physical meaning because the effects of the
quantum mechanics are contained as in the Weyl symbols, so in the
Wigner function, and the last one has no limit at
$\hbar\rightarrow 0$ in the general case \cite{groot}.

Let us consider the Weyl transformation of Eqs. (\ref{syst}) and
expand the Weyl symmetrized product of operators up to
second-order in $\hbar$. This results in:
\begin{eqnarray}
\fl\partial^2_\tau
\alpha+\frac{3}{2}\biggl((\partial_\tau\alpha)^2
+\frac{\hbar^2}{4}(\partial_{k}\partial_{\phi}\partial_\tau\alpha)^2
-
\frac{\hbar^2}{4}(\partial^2_{\phi}\partial_\tau\alpha)(\partial^2_{k}\partial_\tau\alpha)
\biggr) +\frac{3}{2}\biggl((\partial_\tau\varphi)^2
+\frac{\hbar^2}{4}(\partial_{k}\partial_{\phi}\partial_\tau\varphi)^2
\nonumber\\
-
\frac{\hbar^2}{4}(\partial^2_{\phi}\partial_\tau\varphi)(\partial^2_{k}\partial_\tau\varphi)\biggr)-
\frac{3}{2}g\biggl(\varphi^2
+\frac{\hbar^2}{4}(\partial_{k}\partial_{\phi}\varphi)^2 -
\frac{\hbar^2}{4}(\partial^2_{\phi}\varphi)(\partial^2_{k}\varphi)
\biggr)=0,\nonumber\\
\fl\partial^2_\tau \varphi+3\biggl(
\partial_\tau\alpha\partial_\tau\varphi+ \frac{\hbar^2}{4}
(\partial_k\partial_\phi\partial_\tau\alpha)(\partial_{k}\partial_\phi\partial_\tau\varphi)
-\frac{\hbar^2}{8}(\partial^2_{k}\partial_\tau\alpha)
(\partial^2_{\phi}\partial_\tau\varphi)\nonumber\\
~~~~~~~~~~~~~~~~~~~~~~~~~~~~~~~~~~~~~
-\frac{\hbar^2}{8}(\partial^2_{\phi}\partial_\tau\alpha)(\partial^2_{k}\partial_\tau\varphi)\biggr)+
g\varphi=0. \label{num}
 \end{eqnarray}

\noindent where $\alpha(k,\phi,\tau)$ and $\varphi(k,\phi,\tau)$
are the Weyl symbols of the operators $\hat \alpha(\tau)$ and
$\hat \phi(\tau)$, respectively. These equations have to be solved
with the initial conditions at $\tau=0$:
\begin{eqnarray} \nonumber
  \alpha \left(k,\phi, 0 \right) = \ln \left( a \right),~~\partial _\tau  \alpha \left(k,\phi, 0 \right)
   ={\mathcal W}\left[ \sqrt{ -{\frac{{1 }}
{{a^6 }}\frac{\partial^2}{\partial \phi^2} + g\phi ^2 }}\right] , \hfill \\
  \varphi \left(k,\tau, 0 \right) = \phi ,~~~~\partial _\tau  \varphi \left(k,\tau, 0 \right) = \frac{k}
{{a^3 }}. \hfill
\end{eqnarray}

Weyl symbol of the square root \cite{root} can be expressed as
\begin{eqnarray}
{\mathcal W}\left[ \sqrt{ -{\frac{{1 }} {{a^6
}}\frac{\partial^2}{\partial \phi^2} + g\phi ^2
}}\right]=\frac{g^{1/4}}{\pi^{1/2}a^{3/2}}\int_0^\infty
t^{-1/2}\exp\left(-\frac{g\,a^6\phi^2+k^2}{\sqrt{g}\,a^3}\tanh(t)\right)
\nonumber\\
\times \mbox{sech}(t)\left(\frac{g\,a^6\phi^2+k^2}{\sqrt{g}\,a^3}
\mbox{sech}(t)^2+\tanh(t)\right)dt,
\end{eqnarray}
but since the mean values are calculated in the limit
$a\rightarrow 0$, we can take simply $\partial _\tau \alpha
\left(k,\phi, 0 \right)
   =\frac{|k|}{a^3}$ (this will not change the mean values).

  State of the Universe in this
approach is described by the Wigner function $\wp(k, \phi)$, which
is constructed on the basis of definition (\ref{nmm}). That gives:
\begin{eqnarray}\label{wigner} \nonumber
  \wp \left( {k,\phi } \right) = ia\int {\left[ {\left| { - \frac{{\partial ^2 }}
{{\partial \phi ^2 }}} \right|^{ - \frac{1} {4}} \psi ^* \left(
{\phi  + \frac{u} {2}} \right)} \right]\left[ {\left| { -
\frac{{\partial ^2 }} {{\partial \phi ^2 }}} \right|^{ \frac{1}
{4}} \frac{{\partial \psi \left( {\phi  - \frac{u} {2}} \right)}}
{{\partial a}}} \right]e^{iku} du}  -  \hfill \\
  ia\int {\left[ {\left| { - \frac{{\partial ^2 }}
{{\partial \phi ^2 }}} \right|^{ \frac{1} {4}} \frac{{\partial
\psi ^* \left( {\phi  + \frac{u} {2}} \right)}} {{\partial a}}}
\right]\left[ {\left| { - \frac{{\partial ^2 }} {{\partial \phi ^2
}}} \right|^{ - \frac{1} {4}} \psi \left( {\phi  - \frac{u} {2}}
\right)} \right]e^{iku} du,}  \hfill
\end{eqnarray}
or in the momentum representation of the wave function
corresponding to Eq. (\ref{ff}):
\begin{equation}
\wp(k,\phi)=\frac{1}{\pi}\int
c^*(2k-q)c(q)a^{-i|q|+i|2k-q|}e^{2i(q-k)\phi}dq.
\end{equation}

As a result of $a\rightarrow 0$, both Weyl symbols and Wigner
function diverge. In particular, when $a\rightarrow 0$ the Wigner
function becomes more and more oscillating. However the
divergences cancel each other in
  the  expectation values
constructed in the ordinary way. For instance, expectation values
of $\alpha$ and its square are: \begin{eqnarray*} \left\langle
\alpha(\tau) \right\rangle = \int {dkd\phi
\,\alpha(k,\phi,\tau)\wp
\left( {k,\phi } \right)}|_{a\rightarrow 0}\,,\\
 \left\langle
\alpha^2(\tau) \right\rangle = \int dkd\phi\left( \alpha^2
+\frac{\hbar^2}{4}(\partial_{k}\partial_{\phi}\alpha)^2 -
\frac{\hbar^2}{4}(\partial^2_{\phi}\alpha)(\partial^2_{k}\alpha)
\right)\wp ( {k,\phi } )\biggr|_{a\rightarrow 0}.
\end{eqnarray*}

\begin{figure}[h]
\begin{center}
\epsfbox{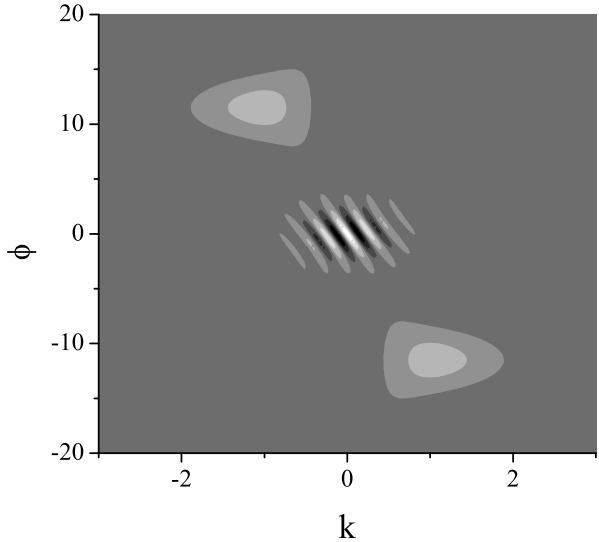}
\end{center}
\caption{\label{fig3} Contour-plot of the Wigner function of
universe at $a=10^{-5}$.}
\end{figure}

An example of the Wigner function for the wave package with
$c\left( k \right) = \frac{4} {{\sqrt 3 }}\left( {\frac{2} {\pi}}
\right)^{\frac{1} {4}} k^2 \exp \left( { - k^2 } \right) $
 is shown in Fig. \ref{fig3}.

\begin{figure*}
\includegraphics[width=17 cm]{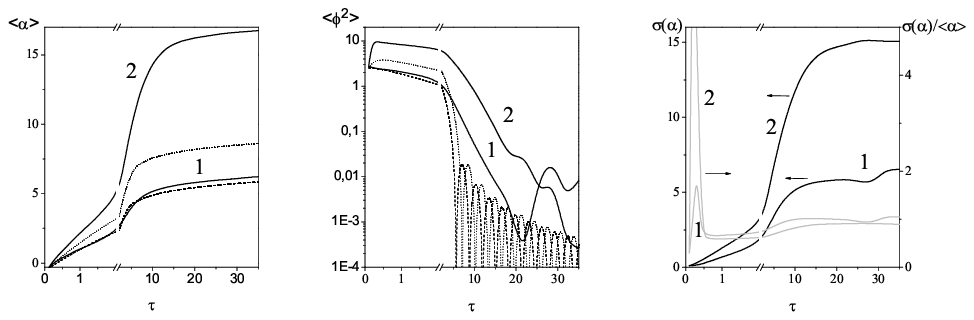} \caption{Evolution of $
\left\langle \alpha  \right\rangle ,\,\left\langle {\varphi ^2 }
\right\rangle$, dispersion $\sigma \left( \alpha  \right) = \sqrt
{\left\langle {\alpha ^2 } \right\rangle  - \left\langle \alpha
\right\rangle ^2 }$ (black curves) and relative dispersion $\sigma
\left( \alpha \right)/\left\langle \alpha  \right\rangle$ (gray
curves) for $c\left( k \right) = \frac{4} {{\sqrt 3 }}\left(
{\frac{2} {\pi}} \right)^{\frac{1} {4}} k^2 \exp \left( { - k^2 }
\right)$. $\hbar = 0$ (1), 1 (2). For comparison we give the
classical evolution with the same initial conditions (initial $k$
is 0 (dashed) and $\sqrt {\left\langle {k^2 } \right\rangle }$
(dotted), initial $\phi =$$\sqrt {\left\langle {\phi^2 }
\right\rangle }$). $a = 10^{-4}$.} \label{fig4}
\end{figure*}

\begin{figure*}
\includegraphics[width=17 cm]{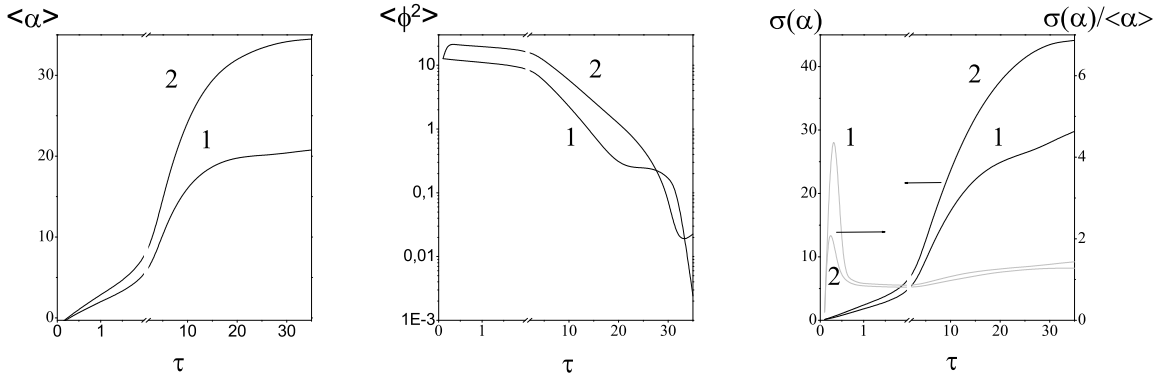} \caption{As in Fig. \ref{fig4} but
for $c\left( k \right) = \frac{20} {{\sqrt 3 }}\left( {\frac{10}
{\pi}} \right)^{\frac{1} {4}} k^2 \exp \left( { - 5 k^2 }
\right)$.} \label{fig5}
\end{figure*}

\begin{figure*}
\includegraphics[width=17 cm]{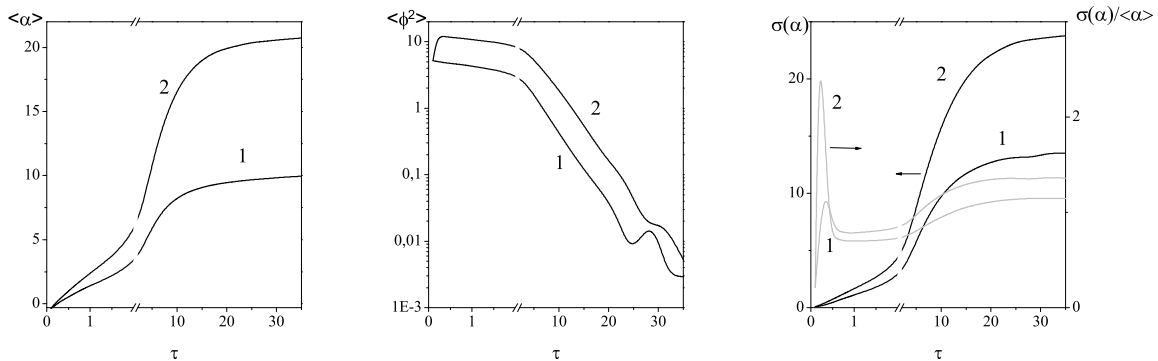} \caption{As in Fig. \ref{fig4} but
for $c\left( k \right) = e^2 \left( {\frac{2} {\pi}}
\right)^{\frac{1} {4}} \exp \left( { - k^2 - 1/k^2} \right)$.}
\label{fig6}
\end{figure*}

As a result of numerical solution of Eqs. (\ref{num}), we have
obtained the evolution of the operators expectation values and
their dispersions. Some results are shown in Figs.
(\ref{fig4}-\ref{fig6}). We have considered three wave packages
providing suppression of the infrared (and certainly ultraviolet)
divergence. For our rather illustrative calculations, the "cuts
off" of the wave packets are chosen of order unity i.e. at the
Plankian level. In principal, the "cuts off" have to be consistent
with the sub-Planskian physics eliminating divergences at
fundamental level. In other words we describe Planskian physis
here, but all sub-Planskian physics should be contained in the
functions of $c(k)$.

It is of interest to compare the evolution numerically obtained
for the arbitrary coupling parameter $g$ with that obtained from
the analytical operator evaluation in the first-order expansion on
$g$ (see previous Section). We consider two different cases for
Eqs. (\ref{num}): i) $\hbar = 0$ (quasi-classical evolution) and
ii) $\hbar = 1$ (quantum corrections; note, that in the previous
sections we didn't use any expansion on $\hbar$).

In addition to results of previous Section, we can see the
deceleration of the initial exponential expansion of the universe
and the transition to the post-inflationary scenario. In the
quasi-classical case the evolution of expectations can be
considered as a good approximation to the classical scenario
(curves 1 and dashed (or dotted) curves, respectively). However,
the inflaton oscillations are smeared in the quasi-classical case
due to presence of the modes with different $k$. Choice of the
more "cold" wave package, i.e. the package with the suppressed
high-frequency modes, enhances the inflation (Fig. \ref{fig5}).
Some enhancement of inflation results also from the removing of
the low-frequency components from the wave package (Fig.
\ref{fig6}).

The quantum corrections (second-order in $\hbar$) essentially
affect the evolution at the initial stage. As it was demonstrated
in previous Section, there is the stage when the scalar field
rolls away the minimum of potential due to kinetic energy of the
wave package (some analog exists also in the classical case, see
dotted curves in Fig. \ref{fig4}). As a result, the inflation
intensifies. However, in any case the inflation comes to the end.

It is important, that in all cases the relative dispersion reaches
some maximum and then decreases. However, the relative dispersion
does not vanish but approaches some asymptotical value, which does
not differ essentially for the different wave packages. It is
astonishing that the asymptotical relative dispersion is similar
in quasi-classical and in quantum-corrected cases. This suggests
that some mechanism of the classical world appearance is required.
Such mechanism is absent in our simplest model. We surmise that
this can be some decoherence produced by the additional degrees of
freedom.

\section{Conclusion}

Quantum evolution of the Universe originated from the some
fluctuation of the scalar field (wave packet) has been considered.

Re-quantization procedure for the Heisenberg operators has been
introduced to compensate the loss of hermicity arising from the
unboundedness of the Universe wave function along $a$ variable.

Mean values of the operators corresponding to the observables,
which are singular at $\tau\rightarrow 0$ on a classical level,
remains singular also in a quantum case.

If the scale like the mass of the scalar field is not introduced,
the evolution of Universe remains quantum eternally in the sense
that the relative dispersion of the scale factor does not decrease
with the increasing $\tau$.

We have solved the equations for the quasi-Heisenberg operators in
the first order on $g$ for the inflation potential
$V(\phi)=g\frac{\phi^2}{2}$. In the first order on $g$ the
dispersion of the logarithm of the scale factor increases with the
increasing $\tau$.

Numerical calculations on the basis of the Weyl-Wigner phase-space
formalism have demonstrated the exit from the inflation with the
approaching of the relative dispersion to some constant but
essentially non-zero value. This value does not depend noticeably
on the universe's wave package.

We are forced to establish that we fail to come to "classical"
Universe (i.e. the Universe with small dispersions) from the
quantum state possing large dispersions. Certainly we can take the
state possessing small dispersion and thus obtain the "classical"
Universe, but it is not so interesting.  More interesting is to
find some peculiar mechanism of the classical world appearance.

\appendix
\section{Solution of the equations for the quasi-Heisenberg operators in the first order on $g$}
 Here we listed the solution of the operator equations for
the quasi-Heisenberg operators in the first order on $g$.
Evaluation is done in the momentum representation of the $\phi$
variable, where the initial conditions have the form: $\hat
\alpha(0)=ln \,a$, $\hat {\dot \phi}(0)=\frac{k}{a^3}$,  $\hat
\phi(0)\equiv \hat \phi= \left(i\frac{\partial}{\partial
k}\right)$, $ \dot{\hat \alpha}(0)=\sqrt{g\hat \phi
^2+\frac{k^2}{a^6}}=\sqrt{g\left(i\frac{\partial}{\partial k
}\right)^2+\frac{k^2}{a^6}} $. In the first order on $g$ the
square root can be extracted giving
\begin{equation}
\dot{\hat\alpha}(0)\approx\frac{|k|}{a^3}+\frac{g}{8}a^3\left(\frac{1}{|k|}\hat\phi^2
+2\hat\phi\frac{1}{|k|}\hat\phi+\phi^2\frac{1}{|k|}\right).
\end{equation}
The solutions of (\ref{syst}) have been obtained in the framework
of the perturbation theory using a computer algebra and have the
form:
\begin{eqnarray}
\fl\hat \phi(\tau)=\hat
\phi+\frac{k}{3|k|}\ln\left(\frac{a^3+3|k|\tau}{a^3}\right)+g\left(\frac{F_2\,\hat
\phi^2+2\hat\phi F_2\,\hat \phi+\hat \phi^2
\,F_2}{4}+\frac{F_1\,\hat \phi+\hat \phi\, F_1}{2}+F_0\right)
\nonumber\\\fl \hat
\alpha(\tau)=\frac{1}{3}\ln\left(a^3+3|k|\tau\right)+g\left(\frac{D_2\,\hat
\phi^2+2\hat\phi D_2\,\hat \phi+\hat \phi^2
\,D_2}{4}+\frac{D_1\,\hat \phi+\hat \phi\, D_1}{2}+D_0\right),
\end{eqnarray}
where functions $F_2,D_2,F_1, \dots$ are

\footnotesize
\[
\fl F_2(k,\tau,a)=\frac{3\,a^6\,{\sqrt{k^2}}\,{\tau }^2 -
6\,a^3\,k^2\,{\tau }^3 - 9\,{\left( k^2 \right)
}^{\frac{3}{2}}\,{\tau }^4}
  {-4\,a^6\,k + 36\,k^3\,{\tau }^2};
\]
\begin{eqnarray*}
\fl F_1(k,\tau,a)=\frac{1}{-108\,a^6\,k^3 + 972\,k^5\,{\tau
}^2}\biggl(-6\,a^9\,k\,{\sqrt{k^2}}\,\tau  + 27\,a^6\,k^3\,{\tau
}^2 - 18\,a^3\,k^3\,{\sqrt{k^2}}\,{\tau }^3 - 27\,k^5\,{\tau }^4
\\\fl-
  2\,a^6\,k\,\left( a^6 - 9\,k^2\,{\tau }^2 \right) \,\ln (a^6) +
  2\,a^6\,{\sqrt{k^2}}\,\left( -a^6 + 9\,k^2\,{\tau }^2 \right) \,\ln (1 - \frac{3\,k\,\tau }{a^3}) +
  2\,a^{12}\,{\sqrt{k^2}}\,\ln (1 + \frac{3\,k\,\tau }{a^3}) \\\fl-
  18\,a^6\,{\left( k^2 \right) }^{\frac{3}{2}}\,{\tau }^2\,\ln (1 + \frac{3\,k\,\tau }{a^3}) -
  2\,a^{12}\,k\,\ln (1 + \frac{3\,{\sqrt{k^2}}\,\tau }{a^3}) +
  12\,a^9\,k\,{\sqrt{k^2}}\,\tau \,\ln (1 + \frac{3\,{\sqrt{k^2}}\,\tau }{a^3})\\\fl +
  36\,a^6\,k^3\,{\tau }^2\,\ln (1 + \frac{3\,{\sqrt{k^2}}\,\tau }{a^3}) -
  108\,a^3\,k^3\,{\sqrt{k^2}}\,{\tau }^3\,\ln (1 + \frac{3\,{\sqrt{k^2}}\,\tau }{a^3}) -
  162\,k^5\,{\tau }^4\,\ln (1 + \frac{3\,{\sqrt{k^2}}\,\tau }{a^3}) \\\fl+ 2\,a^{12}\,k\,\ln (a^6 - 9\,k^2\,{\tau }^2) -
  18\,a^6\,k^3\,{\tau }^2\,\ln (a^6 - 9\,k^2\,{\tau }^2)\biggr);
\end{eqnarray*}
\begin{eqnarray*}
\fl F_0(k,\tau,a)=\frac{1}{1944\,k^3\,\left( -a^6 + 9\,k^2\,{\tau
}^2 \right) }\biggr(-6\,a^9\,k^2\,\tau  - 81\,a^6\,{\left( k^2
\right) }^{\frac{3}{2}}\,{\tau }^2 + 198\,a^3\,k^4\,{\tau }^3+
  297\,k^4\,{\sqrt{k^2}}\,{\tau }^4 + \bigl( -8\,a^{12}\,{\sqrt{k^2}}
   \\\fl+ 72\,a^6\,{\left( k^2 \right) }^{\frac{3}{2}}\,{\tau }^2
     \bigr) \,\ln (a^6) + \left( -8\,a^{12}\,k + 72\,a^6\,k^3\,{\tau }^2 \right) \,\ln (1 - \frac{3\,k\,\tau }{a^3}) +
  \left( 8\,a^{12}\,k - 72\,a^6\,k^3\,{\tau }^2 \right) \,\ln (1 + \frac{3\,k\,\tau }{a^3}) +
  \\\fl\bigl( -14\,a^{12}\,{\sqrt{k^2}} + 12\,a^9\,k^2\,\tau  + 144\,a^6\,{\left( k^2 \right) }^{\frac{3}{2}}\,{\tau }^2 -
     108\,a^3\,k^4\,{\tau }^3
      - 162\,k^4\,{\sqrt{k^2}}\,{\tau }^4 \biggr) \,\ln (1 + \frac{3\,{\sqrt{k^2}}\,\tau }{a^3}) -
  6\,\bigl( -a^6 + 9\,k^2\,{\tau }^2 \bigr) \\\fl\times\bigl( a^6\,{\sqrt{k^2}} + 6\,a^3\,k^2\,\tau  +
     9\,{\left( k^2 \right) }^{\frac{3}{2}}\,{\tau }^2 \bigr) \,{\ln (1 + \frac{3\,{\sqrt{k^2}}\,\tau }{a^3})}^2 +
  8\,a^{12}\,{\sqrt{k^2}}\,\ln (a^6 - 9\,k^2\,{\tau }^2) -
  72\,a^6\,{\left( k^2 \right) }^{\frac{3}{2}}\,{\tau }^2\,\ln (a^6 - 9\,k^2\,{\tau
  }^2)\biggl)\\\fl+\frac{a^3}{216\,k^6}\biggr(-2\,a^3\,k\,{\sqrt{k^2}} - 18\,k^3\,\tau
   - \frac{a^{12}\,k^2}{{\left( a^3 - 3\,k\,\tau  \right) }^3} +
  \frac{a^{12}\,k\,{\sqrt{k^2}}}{{\left( a^3 - 3\,k\,\tau  \right) }^3} +
  \frac{3\,a^9\,k^2}{{\left( a^3 - 3\,k\,\tau  \right) }^2} -
  \frac{3\,a^9\,k\,{\sqrt{k^2}}}{{\left( a^3 - 3\,k\,\tau  \right) }^2} -
  6\,a^{6 - \frac{3\,{\sqrt{k^2}}}{k}}\,k^2
  \\\fl
  \times{\left( a^3 - 3\,k\,\tau  \right) }^{-1 + \frac{{\sqrt{k^2}}}{k}} +
  3\,a^{6 - \frac{3\,{\sqrt{k^2}}}{k}}\,k\,{\sqrt{k^2}}\,{\left( a^3 - 3\,k\,\tau  \right) }^{-1 + \frac{{\sqrt{k^2}}}{k}} +
  18\,a^{3 - \frac{3\,{\sqrt{k^2}}}{k}}\,k^3\,\tau \,{\left( a^3 - 3\,k\,\tau  \right) }^{-1 + \frac{{\sqrt{k^2}}}{k}} +
  \frac{a^{12}\,k^2}{{\left( a^3 + 3\,k\,\tau  \right) }^3}
  \\\fl+
  \frac{a^{12}\,k\,{\sqrt{k^2}}}{{\left( a^3 + 3\,k\,\tau  \right) }^3} -
  \frac{3\,a^9\,k^2}{{\left( a^3 + 3\,k\,\tau  \right) }^2} -
  \frac{3\,a^9\,k\,{\sqrt{k^2}}}{{\left( a^3 + 3\,k\,\tau  \right) }^2} +
  \frac{6\,a^{6 + \frac{3\,{\sqrt{k^2}}}{k}}\,k^2}{{\left( a^3 + 3\,k\,\tau  \right) }^{\frac{k + {\sqrt{k^2}}}{k}}} +
  \frac{3\,a^{6 + \frac{3\,{\sqrt{k^2}}}{k}}\,k\,{\sqrt{k^2}}}{{\left( a^3 + 3\,k\,\tau  \right) }^{\frac{k + {\sqrt{k^2}}}{k}}}
 +
  \frac{18\,a^{\frac{3\,\left( k + {\sqrt{k^2}} \right) }{k}}\,k^3\,\tau }
   {{\left( a^3 + 3\,k\,\tau  \right) }^{\frac{k + {\sqrt{k^2}}}{k}}} \\\fl+
  \frac{9\,a^{\frac{3\,\left( k + {\sqrt{k^2}} \right) }{k}}\,{\left( k^2 \right) }^{\frac{3}{2}}\,\tau }
   {{\left( a^3 + 3\,k\,\tau  \right) }^{\frac{k + {\sqrt{k^2}}}{k}}} -
  9\,{\left( k^2 \right) }^{\frac{3}{2}}\,\tau \,{\left( 1 - \frac{3\,k\,\tau }{a^3} \right) }^{-1 + \frac{{\sqrt{k^2}}}{k}}
   -
  18\,a^3\,k\,{\sqrt{k^2}}\,\ln (a)
 + 3\,a^3\,k\,\left( -k + {\sqrt{k^2}} \right) \,\ln (a^3 - 3\,k\,\tau ) \\\fl+
  3\,a^3\,k^2\,\ln (a^3 + 3\,k\,\tau ) + 3\,a^3\,k\,{\sqrt{k^2}}\,\ln (a^3 + 3\,k\,\tau
  )\biggl);
\end{eqnarray*}
\[
\fl D_2(k,\tau,a)=\frac{a^9\,{\sqrt{k^2}}\,\tau  - 6\,a^3\,{\left(
k^2 \right) }^{\frac{3}{2}}\,{\tau }^3 - 9\,k^4\,{\tau }^4}
  {2\,a^6\,k^2 - 18\,k^4\,{\tau }^2};
\]
\begin{eqnarray*}
\fl D_1(k,\tau,a)=\frac{1}{972\,k^3\,\left( -a^6 + 9\,k^2\,{\tau
}^2 \right) }\biggl(108\,a^9\,k^2\,\tau  + 486\,a^6\,{\left( k^2
\right) }^{\frac{3}{2}}\,{\tau }^2 - 1620\,a^3\,k^4\,{\tau }^3 -
  2430\,k^4\,{\sqrt{k^2}}\,{\tau }^4 \\\fl- 36\,a^{12}\,{\sqrt{k^2}}\,\ln (1 + \frac{3\,{\sqrt{k^2}}\,\tau }{a^3}) -
  216\,a^9\,k^2\,\tau \,\ln (1 + \frac{3\,{\sqrt{k^2}}\,\tau }{a^3}) +
  1944\,a^3\,k^4\,{\tau }^3\,\ln (1 + \frac{3\,{\sqrt{k^2}}\,\tau }{a^3}) \\\fl+
  2916\,k^4\,{\sqrt{k^2}}\,{\tau }^4\,\ln (1 + \frac{3\,{\sqrt{k^2}}\,\tau
  }{a^3})\biggr);
\end{eqnarray*}

\begin{eqnarray*}
\fl D_0(k,\tau,a)=\frac{1}{972\,k^3\,\left( -a^6 + 9\,k^2\,{\tau
}^2 \right) }\biggl(-30\,a^9\,k\,{\sqrt{k^2}}\,\tau  -
81\,a^6\,k^3\,{\tau }^2 + 342\,a^3\,k^3\,{\sqrt{k^2}}\,{\tau }^3 +
513\,k^5\,{\tau }^4 \\\fl+
  10\,a^{12}\,k\,\ln (1 + \frac{3\,{\sqrt{k^2}}\,\tau }{a^3}) +
  60\,a^9\,k\,{\sqrt{k^2}}\,\tau \,\ln (1 + \frac{3\,{\sqrt{k^2}}\,\tau }{a^3}) -
  540\,a^3\,k^3\,{\sqrt{k^2}}\,{\tau }^3\,\ln (1 + \frac{3\,{\sqrt{k^2}}\,\tau }{a^3}) \\\fl-
  810\,k^5\,{\tau }^4\,\ln (1 + \frac{3\,{\sqrt{k^2}}\,\tau }{a^3}) -
  6\,a^{12}\,k\,{\ln (1 + \frac{3\,{\sqrt{k^2}}\,\tau }{a^3})}^2 -
  36\,a^9\,k\,{\sqrt{k^2}}\,\tau \,{\ln (1 + \frac{3\,{\sqrt{k^2}}\,\tau }{a^3})}^2 +
  \\\fl 324\,a^3\,k^3\,{\sqrt{k^2}}\,{\tau }^3\,{\ln (1 + \frac{3\,{\sqrt{k^2}}\,\tau }{a^3})}^2 +
  486\,k^5\,{\tau }^4\,{\ln (1 + \frac{3\,{\sqrt{k^2}}\,\tau
  }{a^3})}^2\biggr)+\frac{a^6}{216\,k^6}\biggr(-8\,k^2 \\\fl+ \frac{a^9\,k^2}{{\left( a^3 - 3\,k\,\tau  \right) }^3} -
  \frac{a^9\,k\,{\sqrt{k^2}}}{{\left( a^3 - 3\,k\,\tau  \right) }^3} - \frac{3\,a^6\,k^2}{{\left( a^3 - 3\,k\,\tau
   \right) }^2} +
  \frac{3\,a^6\,k\,{\sqrt{k^2}}}{{\left( a^3 - 3\,k\,\tau  \right) }^2} +
  6\,a^{3 - \frac{3\,{\sqrt{k^2}}}{k}}\,k^2\,{\left( a^3 - 3\,k\,\tau  \right) }^{-1 + \frac{{\sqrt{k^2}}}{k}} \\\fl-
 \frac{18\,k^3\,\tau \,{\left( a^3 - 3\,k\,\tau  \right) }^{-1 + \frac{{\sqrt{k^2}}}{k}}}{a^{\frac{3\,{\sqrt{k^2}}}{k}}} +
  \frac{9\,{\left( k^2 \right) }^{\frac{3}{2}}\,\tau \,{\left( a^3 - 3\,k\,\tau  \right) }^{-1 + \frac{{\sqrt{k^2}}}{k}}}
   {a^{\frac{3\,{\sqrt{k^2}}}{k}}} + \frac{a^9\,k^2}{{\left( a^3 + 3\,k\,\tau  \right) }^3} +
  \frac{a^9\,k\,{\sqrt{k^2}}}{{\left( a^3 + 3\,k\,\tau  \right) }^3}\\ \fl
  - \frac{3\,a^6\,k^2}{{\left( a^3 + 3\,k\,\tau  \right) }^2} -
  \frac{3\,a^6\,k\,{\sqrt{k^2}}}{{\left( a^3 + 3\,k\,\tau  \right) }^2} +
  \frac{6\,a^{\frac{3\,\left( k + {\sqrt{k^2}} \right) }{k}}\,k^2}
   {{\left( a^3 + 3\,k\,\tau  \right) }^{\frac{k + {\sqrt{k^2}}}{k}}} +
  \frac{3\,a^{\frac{3\,\left( k + {\sqrt{k^2}} \right) }{k}}\,k\,{\sqrt{k^2}}}
   {{\left( a^3 + 3\,k\,\tau  \right) }^{\frac{k + {\sqrt{k^2}}}{k}}}+
  \frac{18\,a^{\frac{3\,{\sqrt{k^2}}}{k}}\,k^3\,\tau }{{\left( a^3 + 3\,k\,\tau  \right) }^{\frac{k + {\sqrt{k^2}}}{k}}}
   \\\fl+
  \frac{9\,a^{\frac{3\,{\sqrt{k^2}}}{k}}\,{\left( k^2 \right) }^{\frac{3}{2}}\,\tau }
   {{\left( a^3 + 3\,k\,\tau  \right) }^{\frac{k + {\sqrt{k^2}}}{k}}} -
  3\,k\,{\sqrt{k^2}}\,{\left( 1 - \frac{3\,k\,\tau }{a^3} \right) }^{-1 + \frac{{\sqrt{k^2}}}{k}} - 18\,k^2\,\ln (a) +
  3\,k\,\left( k - {\sqrt{k^2}} \right) \,\ln (a^3 - 3\,k\,\tau ) \\\fl+ 3\,k^2\,\ln (a^3 + 3\,k\,\tau ) +
  3\,k\,{\sqrt{k^2}}\,\ln (a^3 + 3\,k\,\tau )\biggl).
\end{eqnarray*}
\normalsize
The output has been generated from
\textit{MATHEMATICA}; $\sqrt{k^2}$ means $|k|$.

\end{document}